# The optical near-field of an aperture tip

A. Drezet, M. J. Nasse, S. Huant and J. C. Woehl

Laboratoire de Spectrométrie Physique, Université Joseph Fourier Grenoble et CNRS,

38402 Saint Martin d'Hères, France

**We use fluorescent nanospheres as scalar detectors for the electric-field intensity in order to probe the near-field of an optical tip used in aperture-type near-field scanning optical microscopy (NSOM). Surprisingly, the recorded fluorescence images show two intensity lobes if the sphere diameter is smaller that the aperture diameter, as expected only in the case of vector detectors like single molecules. We present a simple but realistic, analytical model for the electric field created by light emitted by a NSOM tip which is quantitative agreement with the experimental data.**

A crucial step towards the development of nanoscale optoelectronic devices from building blocks like single molecules, semiconductor nanocrystals, or quantum dots is the study of their behavior under optical stimulation. On these length scales, near-field effects and diffraction phenomena play a major role both for the interaction of such subwavelength structures with incident light as well as for the communication pathways between them. While diffraction phenomena are among the most important and intensively studied effects in optics with a wide range of applications in other domains of physics (*1-3*), an analysis of the situation is not always straightforward. Diffraction by an object bigger than the wavelength of the incident electromagnetic radiation can be described using either classical, scalar theory (*4*) or an electromagnetic approach based on Maxwell's equations (*5*). Diffraction by objects (or apertures) comparable to or much smaller than the wavelength of the incident radiation, however, is more difficult to analyze, and only few analytical solutions for specific geometries are known. With the increasing research activity in nano-optics in recent years (*11*), the open question of how light passes through a subwavelength aperture and interacts with the sample has attracted a lot of interest. An answer to the problem is not only of fundamental importance for image interpretation and optical resolution in near-field optics, but also for the detection and addressing of single nano-objects (single molecules (*12-16*), semiconductor nanocrystals (*17*), luminescent centers in quantum well structures (*17a*), and quantum dots (*18,19*)), and for applications in fields like single photon sources (*20,21*) and quantum optics (*19*).

In order to address the question of how light is diffracted by a subwavelength aperture, we use a near-field scanning optical microscope (NSOM) tip in illumination mode (*22-24*) as a model system, i.e. a tapered, optical single mode fiber covered with a thin (100 nm) Al coating presenting a small, circular aperture at the tip apex. The electric field close to the aperture is imaged using fluorescent nanospheres which, depending on their size, yield qualitatively different fluorescence images. A frequently cited model for interpreting such images is the Bouwkamp solution (*25*) for a circular, subwavelength hole in a conducting screen (which we will refer to as a Bethe (*6*) aperture in planar geometry) illuminated by an incident plane wave, which seems to be a rather questionable model for an optical fiber tip (a Bethe aperture in conical geometry). Often, elaborate numerical calculations for specific tip-sample configurations are carried out (*26-28*) which, however, depend on the choice of boundary conditions as well as the discretization and iteration procedures used. It appears therefore important to develop a simple but realistic, analytical description of the tip's electromagnetic field, and especially of its electric field component since it dominates the interaction with fluorescent nano-objects. In previous work, we have given an analytical description of the electromagnetic far-field of a NSOM tip (*29,30*). We now present a simple, analytical model for the electric field component close to the tip which correctly takes into account its conical geometry, and, in contrast to the solution for the planar Bethe aperture, accurately reproduces the experimental results.

It is well known that fluorescent molecules, when in resonance with the excitation light, are selective detectors for the incident light polarization since their fluorescence intensity is proportional to $(\boldsymbol{\mu}\cdot\mathbf{E})^2$ where $\boldsymbol{\mu}$ is the molecular transition dipole moment and $\mathbf{E}$ the electric component of the excitation field. This property has been used to characterize, for example, the squared electric field components in the focus of a high numerical aperture lens (*31*) or near an optical fiber tip (*12,14*). The analysis of the acquired images, however, requires an exact knowledge of the molecular transition dipole moment in all three directions. This is not the case with nanospheres of small diameter which contain a large number of fluorescent molecules that are uniformly distributed throughout the sphere volume. Since their transition dipole moments are randomly oriented, the fluorescence intensity of such an ensemble of $N$ incoherently emitting molecules is proportional to

$$I \propto \sum_{i=1}^{N} (\mathbf{\mu}_i \cdot \mathbf{E}_i)^2 \approx \frac{N}{3} \mu^2 \langle E^2 \rangle_{\text{sphere volume}}$$

where µ is the typical value for the molecular transition moment. Fluorescence labeled nanospheres act therefore as isotropic volume detectors of the average electric field intensity and can be used to produce an intensity map of the electric field without any further knowledge of orientational parameters (*32*). In contrast to single molecules which are *vector* detectors, fluorescent nanospheres, in the limit of very small sizes, act as *scalar* detectors of the electric field. It is clear that details in the electric field distribution of a NSOM tip can only be picked up when the sphere diameter is significantly smaller than that of the optical aperture. This behavior is shown in Fig. 1A et B which presents the recorded fluorescence images for two different ratios of sphere and aperture diameter. When the sphere is bigger than the tip aperture (Fig. 1A), the recorded fluorescence profile is smooth and presents no substructures, while two fluorescence lobes appear in the case where the sphere is smaller than the tip aperture (Fig. 1B). As can be seen, all observed nanosphere fluorescence images have the same orientation and shape which means that the nanospheres probe the optical tip and not vice-versa.

To explain the observed intensity distribution, we have developed a simple model for the electric field around the metal coated aperture tip. Our model is based on the assumption that the electric field $\mathbf{E}\exp(-i\omega t)$ produced by the tip is essentially static (satisfying the potential condition $\nabla \times \mathbf{E} = \mathbf{0}$) and is completely characterized by the electric charge distribution located on the metal coating of the tip apex. If we call $(x,y)$ the aperture plane and $z$ the tip axis (with the incident light polarized along $x$ and propagating in the $z$ direction), we can assume a surface charge density $\sigma$ of the form

$$\sigma(\rho,\phi,z) = f(\rho,z)\cos\phi$$

for any point $[x,y,z] = [\rho,\phi,z]$ on the metal coating, where $\phi$ is the angle with respect to the polarization plane. The $\cos\phi$ dependence is explained by the fact that the linearly polarized, optical $LP_{0,1}$ mode in the single mode fiber is mainly coupled to the fundamental, transverse electric (TE) mode of the conical wave guide preceeding the aperture zone (*33*, *34*). This guided mode will induce polarization charges in the surrounding metal coating at the tip apex showing the same $\cos\phi$ dependence. In the simplest case, we can assume a linear, ring-like charge density

$$\ell(\phi) = \ell_0 \cos\phi$$

concentrated at the rim of the tip aperture of radius $a$ (Fig. 2A). A map of $E^2$ and the squared electric field components is presented in Fig. 3A for the $(x,y)$ plane at $z = 0.3a$ from the aperture plane. This ring model produces two intensity lobes for the total electric field and is therefore able to qualitatively explain the experimentally observed emission lobes from Fig. 1B. Nevertheless, it is not satisfactory because i) it does not obey Maxwell's boundary conditions on the metal, and ii) it implies a logarithmic energy divergence at the aperture rim. This latter point is forbidden by the classical analysis of light diffraction by an edge or a corner (*35*).

In our problem, the edge-corner condition naturally leads to the requirement that the electric surface charge density increases not faster than $(1/r)^n$ where $r$ is the distance from the rim (Fig. 2B) and where $n = 1 - \pi / (2\pi - \beta)$ is a function of the corner angle $\beta$ (*5*). We must therefore impose

$$\sigma = g(\rho, z) / r^n \cos\phi$$

($g$ has no singularity at the rim), and we suppose $g = \sigma_0 =$ const since the variation of $g$ is negligible for objects, like the tip apex, with a size on the order of or smaller than one wavelength. The angle $\beta$ and the tip angle $\alpha$ are related by $\alpha = \pi - 2\beta$. The electric field produced by such a surface charge distribution is represented in Figs. 2B and 3B and shows the same intensity lobes as the simpler ring model. It is interesting to note that the image representing $E^2$ is entirely dominated by the $x$ and $z$ components of the electric field, and carries direct information on their squared maximum values in the image plane: the $x$ component is highest in the image center (where the $z$ component is negligible) while the $z$ component is highest in the lobe centers (where the $x$ component is very weak). A comparison with the Bouwkamp solution for a planar Bethe aperture (Fig. 3C) shows strong, qualitative differences with respect to the presented models. This model is unable to even qualitatively account for the experimentally observed fluorescence images from Fig. 1B.

On the basis of the proposed surface charge model, we can simulate the fluorescence images of nanospheres obtained by NSOM by integrating the electric field intensity in over the sphere volume. Fig. 1C shows such a simulation for a sphere to aperture diameter ratio of 0.37. The image is obtained for an optical tip with an aperture of radius $a = 300$ nm, a 100 nm metal coating and a cone angle α of 30 degrees. The simulation assumes that the tip keeps its vertical position constant while scanning over the sphere. This is in good agreement with the experimental fact that the topographic image does not vary significantly over the small fluorescent region of the nanosphere (*36*). For the simulation, a sphere on the sample surface is considered as an (ideal) volume detector for the electric field intensity as discussed above. As a confirmation of the experimental observations, the emission lobes disappear and merge into one spot when the sphere diameter becomes several times larger than the aperture diameter.

Fig. 4A presents simulated line scans along the symmetry axis of fluorescence images with increasing axial distance $\Delta z$ (*i.e.* the distance between the aperture plane and the closest point on the sphere surface) for fluorescence imaging of the smallest commercially available fluorescent nanospheres with a diameter of 20 nm by a typical NSOM tip (optical aperture of 100 nm in diameter, cone angle α of 30 degrees, 100 nm Al coating). As can be seen, the separation between the two lobes decreases with increasing axial distance. This can be explained by the fact that the two lobes are essentially due to the $E_z$ component which decays rapidly with increasing distance from the aperture rim. It can be noted that this near-field effect is a general feature since both the ring model as well as the surface model give rise to a very similar behavior (see the dotted and solid lines shown in Fig. 4A). Fig. 4B presents the variation of the contrast ratio $I_{max}/I_{center}$ between the maximum intensity on top of the lobes versus the center intensity with increasing axial distance. The dependence is very strong for the first 20 nm but levels out at large distances from the tip aperture. (In this case, the differences between the surface model and the ring model are much more pronounced.) It should be emphasized that this ratio is a direct experimental measure for the squared ratio of the *z* versus *x* components of the electric field (as discussed earlier) in the limit of small nanospheres.

On the basis of the surface model, we propose to use the lobe separation and contrast ratio $I_{max}/I_{center}$ from fluorescence images of small nanospheres to determine *in situ* both the aperture radius *a* and the axial (feedback) distance $\Delta z$ under the actual experimental conditions. In addition, the polarization of the incident light can be deduced from the symmetry axis passing through the lobes. For the experiment presented in Fig. 1B, we have determined the aperture radius and axial distance to be $a = 300$ nm and $\Delta z = 18$ nm. In this case, the experimental profiles of fluorescence intensity along the symmetry axis are in perfect agreement with the simulated curve (see Fig. 1D).

Our models differ strongly from the Bouwkamp solution for the diffraction of a plane wave by a Bethe aperture in planar geometry. As seen in Fig. 3, the $E_z$ component is dominant for the surface and ring models, whereas the Bouwkamp field is dominated by the $E_x$ component. The $E_z$ term never dominates the $E_x$ component even in the context of the complete solution as given by Meixner-Andrejewski (*7*) and to which the Bouwkamp expression is the first order approximation in powers of $k = \omega/c$. A planar Bethe aperture illuminated by a plane wave is not an adequate description for the optical tip because both geometry and incident wave topology are very different from the real situation.

In addition, experimental far field measurements prove the existence of an electric dipole term (*37*) which is present in our models but is completely missing in all solutions for the Bethe aperture in planar geometry. More precisely, the measurements show that there are effective electric ($\mathbf{P}_{eff}$) and magnetic dipoles ($\mathbf{M}_{eff}$) located in the aperture plane which are related by the equation $\mathbf{M}_{eff} = 2\hat{\mathbf{z}} \times \mathbf{P}_{eff}$ (where $\hat{\mathbf{z}}$ denotes the propagation direction in the fiber). A modal analysis based on an expansion in conical transverse electric and magnetic modes justifies these observations only if an electrostatic and magnetostatic field is present in the aperture zone. However, all solutions to the Bethe problem yield only a magnetostatic but no electrostatic (zero order) term. It can be added that our models do not analyze the magnetostatic field of the fiber tip, but that it can be easily obtained in a similar manner. The existence of this component and its influence on the electric field, however, can be safely ignored in the present case which is based on electric dipole transitions.

We have presented a simple, theoretical model for the electric field of light transmitted by a conical NSOM tip. Our model is able to explain, both qualitatively and quantitatively, the recorded fluorescence images of nanospheres that are small compared to the aperture diameter. The widely admitted Bouwkamp solution for the planar Bethe aperture fails to explain these images even qualitatively. Several interesting differences between these models can be noted, in particular a much stronger *z* component in our model as compared to the Bouwkamp solution. We plan to address this distinctive feature by NSOM imaging of fluorescent molecules under polarization detection. Fluorescent nanospheres are valuable objects since they act as isotropic volume detectors for the squared electric field intensity. Together with the proposed model, they can be used for an *in situ* determination of both tip aperture and axial distance under the actual experimental conditions which is significant for practical purposes. We also plan to use these objects in order to establish a detailed electric field intensity map of a NSOM tip in all three dimensions. Based on the proposed model, it is possible to determine the orientation of single molecule emitters in all three dimensions from recorded fluorescence images, which has important implications for the addressing of single molecules and optical nanomanipulations. It is also clear that the electric field distribution at the fiber tip has a direct influence on the excitation properties of single nano-objects located underneath the tip that can be used for potential photonic nanosources and quantum optical experiments in the near-field domain (*19*).

39. Supported by grants from the Volkswagen Foundation (Program "Physics, Chemistry and Biology with Single Molecules"), the Institut de Physique de la Matière Condensée, and the CNRS.

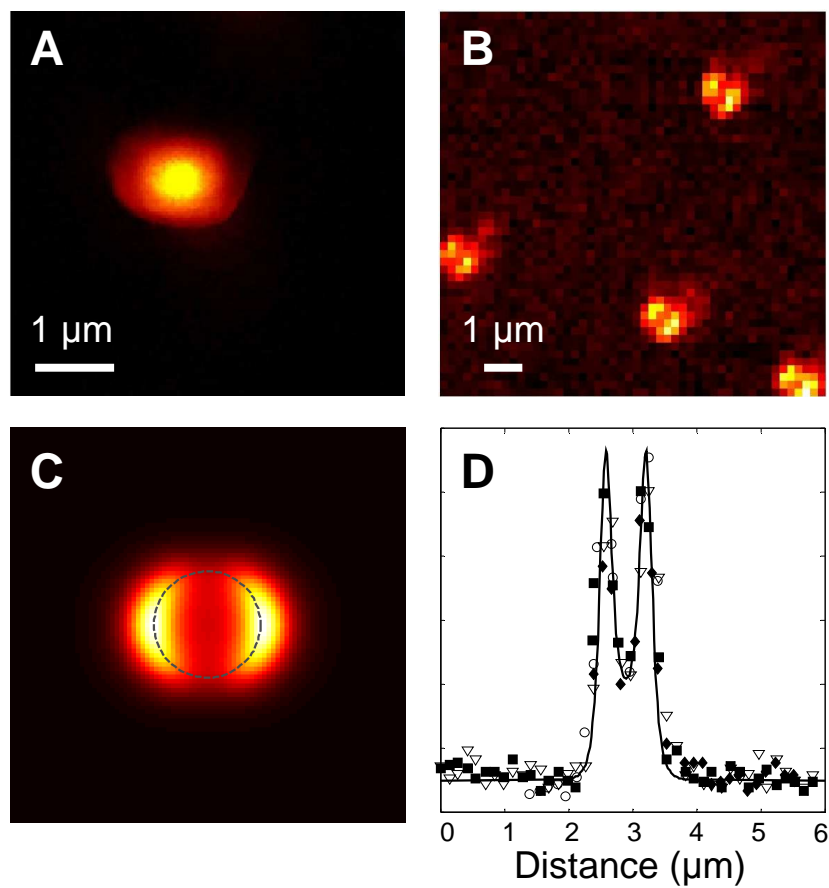

**Fig. 1.** Fluorescence images of nanospheres of different size taken by a NSOM tip. The images show carboxylate-modified, yellow-green fluorescent nanospheres (Molecular Probes) with diameters of (**A**) 500 nm ± 5% and (**B**) 220 nm ± 5% , respectively, which are deposited from basic solution (pH 10-11) on a PMMA layer spin-coated onto a clean glass cover slide. A few hundred μW of the 514.5 nm line of an Argon ion laser are coupled into a single mode optical fiber with a tapered, metal coated end presenting an optical aperture (typical transmission of $10^{-3}$ to $10^{-2}$) which excites the sample, held at nanometer distance using a feedback loop with tuning fork detection (*38*). The emission

is collected in the far field above the sample surface using an Al mirror and, after removal of residual excitation light using color glass filters (Schott) and interference filters, detected by (**A**) a low dark count avalanche photodiode module (EG&G) or (**B**) a channel photomultiplier (PerkinElmer) in photon counting mode. The integration time for each point in the image is 100 ms and 50 ms, respectively. (**C**) Simulated fluorescence image of a 220 nm diameter sphere scanned by a 600 nm diameter aperture tip (indicated by the overlaid circle) at an axial distance of $\Delta z = 18$ nm between the aperture plane and the closest point on the sphere surface. (**D**) Horizontal cross-section of (C) together with fluorescence intensity profiles from the four nanospheres (different symbols) shown in (B). The theoretical and experimental curves are in perfect agreement.

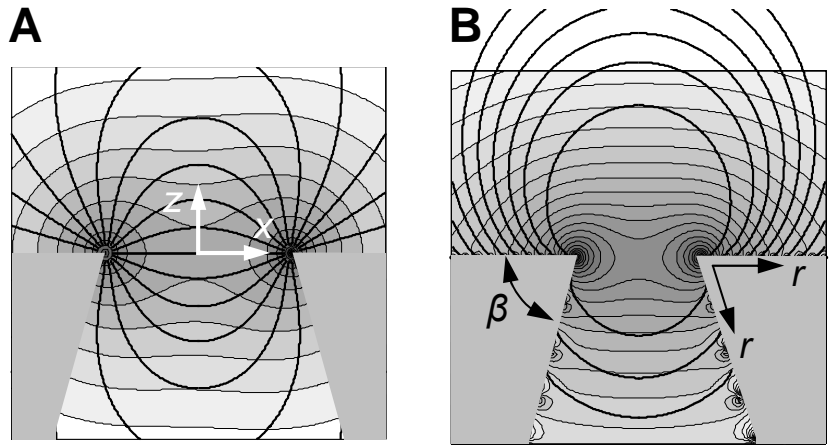

**Fig. 2.** Cross-section of an optical fiber tip with the associated electric field lines and a logarithmic intensity map. The polarization of the incident light propagating in the fiber core (*z* direction) is oriented along the *x* axis. The electric field is generated by polarization charges in a perfect metal coating modeled by (**A**) a linear charge distribution around the aperture rim, and (**B**) a surface charge distribution on the lateral and inner surface of the metal coating surrounding the tapered fiber (aperture diameter 2*a*). The parameters *r* (distance from the aperture rim) and the cone angle *β* are indicated in the figure. The electric field lines in the immediate vicinity of the metal surface show artefacts mainly due to the numerical discretization procedure, which become, however, insignificant at larger distances.

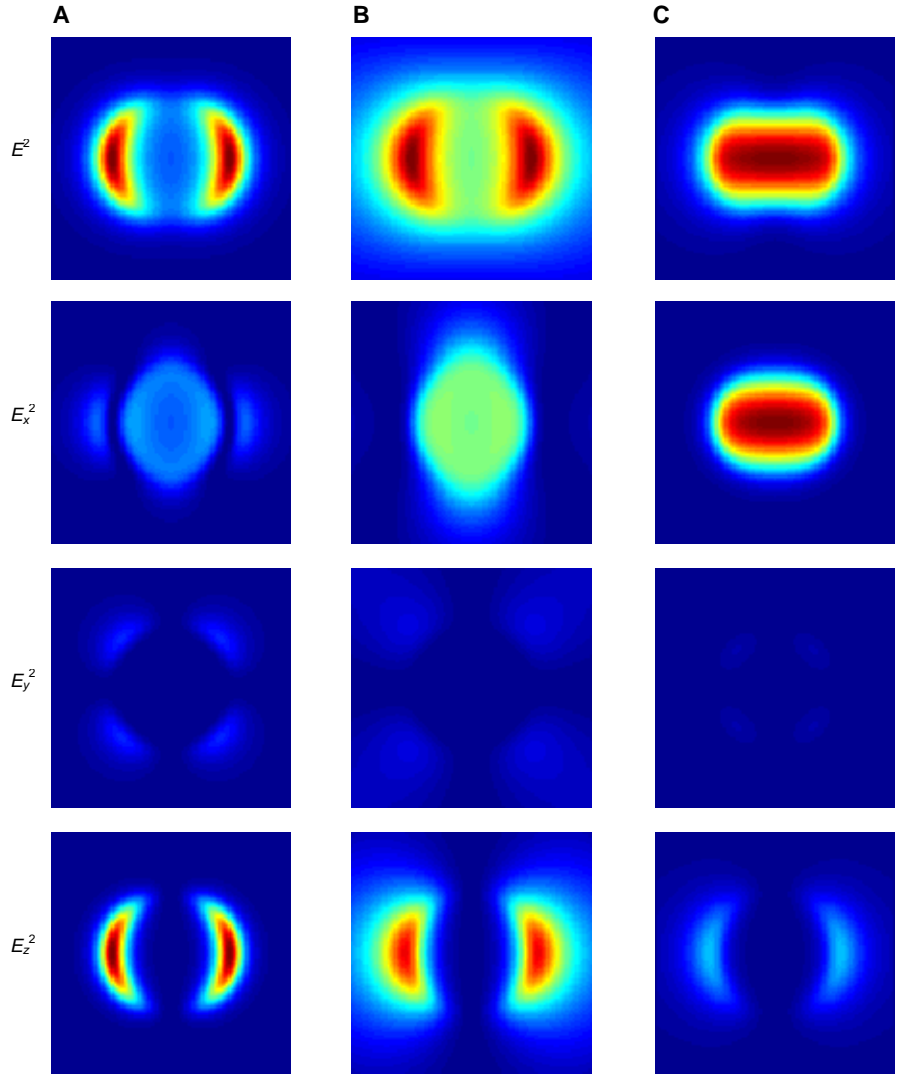

**Fig. 3.** Intensity maps for the electric field ($E^2$, $E_x^2$, $E_y^2$, and $E_z^2$) at a distance of $z = 0.3a$ from the tip aperture of radius $a$ (the images in each column share the same colorbar). Electric field created by (**A**) a ring-like charge distribution on the rim of the optical aperture as illustrated in Fig. 2A, (**B**) a surface charge distribution as illustrated in Fig. 2B, and (**C**) a planar Bethe aperture (Bouwkamp solution). The images on the first line

correspond to those that would be obtained using a pointlike, scalar detector for the electric field (e.g. an idealized, infinitely small fluorescent nanosphere). As can be seen, only the first two models are able to produce two lobes for the $E^2$ distribution. The last three lines correspond to fluorescence images predicted for pointlike vector detectors for the electric field (e.g. single fluorescent molecules) oriented along the $x$, $y$, and $z$ direction, respectively. While the $E^2$ intensity map in the case of the surface model is similar to the image in Fig. 1C, the latter was obtained by averaging the electric field intensity over the sphere volume in all three dimensions for each point of the image (the closest point on the sphere surface is at $z = 0.06a$, and the sphere center at $z = 0.43a$ from the aperture plane).

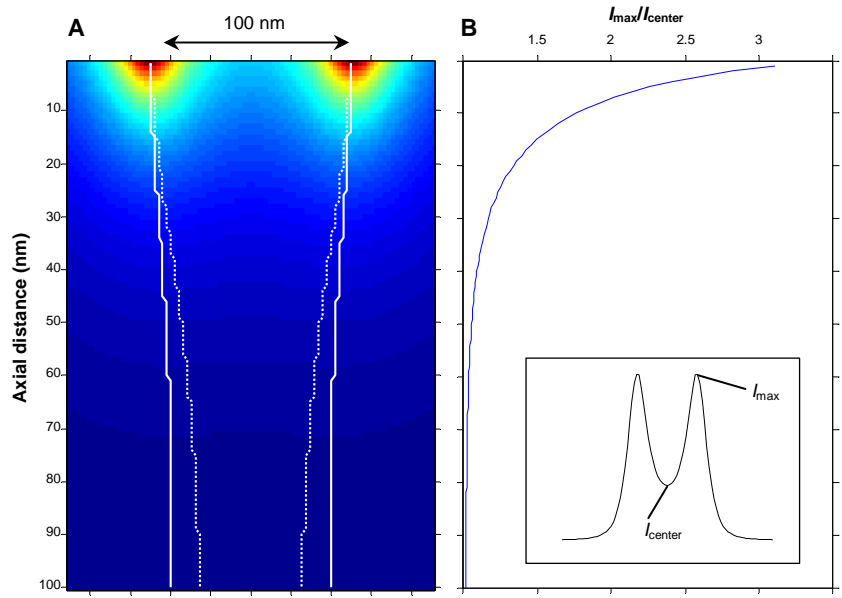

**Fig. 4.** Theoretical evolution of characteristic parameters for fluorescence imaging of nanospheres with increasing axial distance $\Delta z$. (**A**) Cartography of the fluorescence intensity in the (*x,z*) plane and lobe separation distance as a function of $\Delta z$. The solid and dotted lines mark the intensity maxima with increasing axial distance for the surface and ring charge model, respectively. (**B**) Variation of the contrast ratio between maximum intensity on top of the lobes ($I_{max}$) versus the center intensity ($I_{center}$, see insert) as a function of $\Delta z$ (same scale as in **A**).